# Bayesian inference and the world mind

John O. Campbell

## Abstract
Knowledge is a central concept within both Bayesian inference and the mathematical and philosophical program of logic and semiotics initiated by Charles Sanders Peirce and further developed by George Spencer-Brown and Louis Kauffman. The latter school is more philosophical than is usual with the practitioners of Bayesian inference and claims the existence of a world mind. When these two disciplines inform each other semiotics is provided with mathematical mechanism and Bayesian inference is seen to be closely related to the act of distinction, the fundamental basis of logic in the work of Spencer-Brown. Given that Darwinian processes are physical implementations of Bayesian inference and are utilized within numerous scientific theories across a wide range of disciplines as mechanisms for the creation and evolution of knowledge we may view the conjunction of these theories, within universal Darwinism, as descriptive of a world mind.

## Introduction
Bayesian probability is often defined as representing a state of knowledge (Jaynes 1986) and Bayesian inference is the primary mathematical tool for describing knowledge. Another mathematical and philosophical approach to knowledge is found within the work of Charles Sanders Peirce and George Spencer-Brown. Louis Kauffman observed that Spencer-Brown, with the Laws of Form (1979), completed a program to develop the foundations of mathematical logic which had been begun by Peirce and provides a comment on the significance of this program (Kauffman 2001):

> *It remained for Spencer-Brown (some fifty years after Peirce and Nicod) to see the relevance of an arithmetic of forms underlying his notation and thus putting the final touch on a development that, from a broad perspective, looks like the world mind doing its best to remember the significant patterns that join logic, speech and mathematics.*

Peircean semiotics is a related but more philosophical effort to understand 'the world mind'. It describes knowledge as involving a triad of components: the object, a sign which signifies the object, and an interpretant or the mental effect caused upon a person by the sign. It will be of interest that Peirce considered his linking of interpretant to a human mental state as a simplification to ease understanding (Peirce 2001):

> *My insertion of "upon a person" is a sop to Cerberus, because I despair of making my own broader conception understood. [...]*

Presumably this broader conception understands that there are components other than those related to humans which compose the world mind.

Spencer-Brown's mathematics and philosophy start with the concept of a distinction. From this concept he is able to derive mathematical logic, Boolean algebra and thereby the foundations of mathematics. Mathematics includes probability but this route from distinction to probability is circuitous. The demonstration by E.T. Jaynes that logic is entailed within probability theory as a special case (2003) suggests that a shorter route may exist.

Inherent to Spencer-Browns conception of 'distinction' is both a 'mark' to mark the distinction and an observer. The concluding sentence of Laws of Form, is:

> *We see now that the first distinction, the mark and the observer are not only interchangeable, but, in the form, identical.*

It has also been noted that Spencer-Brown's paradigm meshes well with Peircean semiotics (Kauffman 2001) if we equate distinction with object, mark with sign and observer with interpretant. The choice of the word 'observer' should be understood here in the Einsteinium manner as a generalized entity able to react to measurements or data or some other instantiation of a sign. I suggest the word 'model' be used to guard against anthropocentric interpretations.

Although the model of logic and semiotics developed by Peirce and Spencer-Brown may be suggestive of a world mind capable of building and processing knowledge it lacks details. Clearly it offers some insight into the relationship between models (interpretants) and that which is modeled (objects) a relationship we might consider to be knowledge however the mechanisms which relates object to sign and sign to model are left unspecified.

If semiotics is to serve as a useful paradigm for knowledge it should be expected to encompass science. Indeed, science, understood as a process of Bayesian inference, may be cast in a triad form compatible with semiotics: phenomena, data, model. The inclusion of Bayesian inference brings a particular strength to the semiotic paradigm. Bayesian experimental design provides a mechanism for extracting data (signs) from the phenomena (object) so as to maximize the expected knowledge gain of the model (interpretant). It also provides the mechanism for using data to most effectively update models in the form of the Bayesian update.

Probability distributions employed as models within Bayesian inference have the property of information entropy; the amount of information which separates the model from certainty. If the log of the probabilities used in the inference is to base two then the information is measured in bits. Bits are the basic unit of distinction. The model's entropy is the number of distinctions which separate it from certainty. While the models treated within Bayesian inference may be separated from certainty by any finite number of distinctions the special case of one bit separations from certainty is the case where Bayesian inference becomes isomorphic with classical logic.

Models within Bayesian inference are moved towards certainty by incorporating the implications of data (sign) into the model (interpretant); through the process of the Bayesian updating. Mathematically this is the unique method of moving a model towards certainty (Jaynes 2003); that is of increasing knowledge. Bayesians have frequently committed the error of assuming that knowledge is a human property and that Bayesian inference is descriptive of a solely human activity. Spencer-Brown did not indulge in this misconception; he maintained a focus on the world mind.

> *Thus we cannot escape the fact that the world we know is constructed in order (and thus in such a way as to be able) to see itself.*
>
> *This is indeed amazing.*
>
> *Not so much in view of what it sees, although this may appear fantastic enough, but in respect of the fact that it can see at all.*
>
> *But in order to do so, evidently it must first cut itself up into at least one state which sees, and at least one state which is seen. In this severed and mutilated condition, whatever it sees is only partially itself.*

It can be appreciated that this world mind is well described within the context of science if we accept that Bayesian inference is isomorphic to the Darwinian process. The operation of a Darwinian process is a physical implementation of the mathematics of Bayesian inference (J. O. Campbell 2009). Scientific theories which describe the creation and evolution of their subject matter as due to the operation of Darwinian processes are ubiquitous across disciplines within the scientific literature (Campbell, Universal Darwinism; The path of knowledge 2010). Taken as a whole these theories are descriptive of a world mind.

## Bayesian inference and semiotics

Bayesian inference may be understood in conjunction with information theory. The basic relationship is:

$$I(\omega_n) = -log(P(\omega_n))$$

Where $\omega_n$ is the nth possible distinguishable state of event $\omega$. The complete set of probabilities assigned to the $\omega_n$ form a probability distribution which sums to 1. We can consider such $\omega_n$ as equivalent to an exhaustive and mutually exclusive set of hypotheses $h_n$ composing a model H. In this case information is defined:

$$I(h_n) = -log(P(h_n))$$

When the log of the probability is base 2 information is measured in bits which are the basic unit of distinction.

Information measures the surprise we can expect if $h_n$ proves to be true. If the hypothesis is assigned a small probability then we receive a large amount of information if we become certain of its truth and vice versa.

The above definition of information in terms of probability also provides us with a definition of probability in terms of information:

$$P(h_n) = 2^{-I(h_n)}$$

$P(h_n)$ is the measure of plausibility assigned to hypothesis $h_n$ being true and it is equal to $2^{-I(h_n)}$ where $I(h_n)$ is the amount of information received if $h_n$ is found to be certain. This definition makes clear probability's epistemic nature; it is a function of incomplete information.

Such sets of mutually exclusive and exhaustive hypotheses provide a complete model of the outcomes of event $\omega$. The model contains a listing of all possible distinguishable states. Such models have the property of entropy E which is the expected amount of information required to bring the model to certainty. Entropy is measured in units of bits. Thus entropy is the number of binary distinctions which must be made in order for the model to arrive at certainty.

Using the equation defining information we see that there is an inverse to entropy or ignorance; it is a probability which may be used as a definition of knowledge K:

K = 2$^{-E}$

K is the measure of plausibility assigned to a model's ability to predict correct outcomes and it is equal to 2$^{-E}$ where E is the entropy of the model. Knowledge and entropy describe our state of knowledge in relation to certainty. One is merely a re-scaling of the other; entropy measures the number of true-false distinctions which separate the state of a model from certainty, knowledge measures the distance the model has come towards certainty.

A model can only be considered a model if it maintains some accuracy in its representation of the phenomena it is modeling. It is a mathematical theorem that such accuracy may only be maintained by the model if its probabilities are reassigned on the reception of new information in accordance with Bayesian updating.

$$P(h_n|IX) = P(h_n|X)\frac{P(I|h_nX)}{P(I|X)}$$

The concepts we have described: information, probability, models and Bayesian updates are components of what might be called inferential systems. Together with phenomena the inferential system forms a type of semiotic process:

1) Information is a sign which signifies aspects of phenomena
2) A model in the form of a probability distribution is the interpretant of this information when the model is updated to reflect the implications of the information.

Further an inferential system may be understood in terms of a semiotic process based on Spencer-Brown's fundamental concept of distinction. Models assign probabilities to distinguishable states. The current state of a models is reflected in its entropy; the number of distinctions which separate the model from certainty.

Semiotics may also serve to remind us that the inferential process is a unity. Information, probability, models and Bayesian updates are concepts which require each other. For instance light from a distance star may contain information concerning the nature of the intervening spaces through which it has travel if it is able to interact with and update a model, such as one contained in an astronomers brain. However no distinctions are made due to that light if no model is updated with the implications of those distinctions. Light from a star falling on a rocky outcrop contains little information.

## Bayesian inference and Darwinian processes

This author has demonstrated the one to one correspondence between the steps of the algorithms for Bayesian processes and Darwinian processes (2009, 2010). However this correspondence is obvious with even informal observations.

If we examine this process within the familiar context of biology we can identify characteristics with genes and varieties of these characteristics with the various alleles of the gene. The probability distribution of the alleles shifts from generation to generation according to selection pressure. The formula used to describe this shift is:

$$p' = \frac{NpR_A}{NpR_A + NqR_B} = p\frac{R_A}{\bar{R}} \qquad \text{(Ricklefs 1979)}$$

Where p' is the probability of the particular allele in the latter generation, p is the probability of the particular allele in the former generation, $R_A$ is the fitness of the particular allele and $\bar{R}$ is the average fitness of all competing alleles. This is clearly a Bayesian update of the probabilities making up the alleles' distribution.

Another persuasive illustration of the isomorphism between inferential systems and Darwinian processes involves a device designed by Francis Galton (1877) and used in a lecture to the Royal Society. The device, pictured below, has recently been re-discovered within the statistics community and repurposed as a 'visualization of Bayes' theorem' (Stigler 2011).

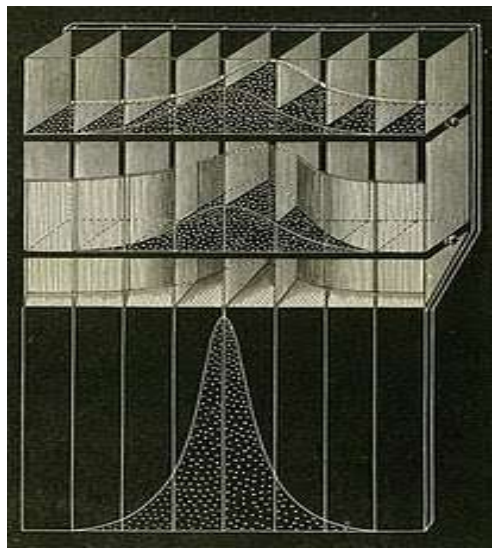

The device contains three compartments: a top one representing the prior distribution, a middle one representing the application of the likelihood to the prior and a third representing the normalization of the resulting distribution. Beads are loaded in the top compartment to represent the prior distribution and then are allowed to fall into the second compartment. The trick is in the second compartment where there is a vertical division in the shape of the likelihood distribution. Some of the beads fall behind this division and are 'wasted'; they are removed from sight and the remaining beads represent the product of the prior and likelihood functions, the 'Bayesian update'.

Perhaps the most remarkable thing about this demonstration is that Galton never mentioned Bayes in his lecture; he had probably never heard of him or his theorem. Rather, his purpose in building the device was to demonstrate the principle of natural selection. The top compartment represents the child generation with variations and the middle compartment represents the individuals selected from that generation.

That this device may be used to model both Bayesian inference and a Darwinian process illustrates an isomorphism between the two. The pertinent implication is that any system driven by a Darwinian process is performing inference. The predictive accuracy of its internal model is increased by processing the data available to it, its experiences in the broader world.

## Conclusion

The work of Spencer Brown is provocative and its reception varied but muted. Extreme claims have been made for its mathematical importance. The book jacket of some editions of Laws of Form contains an endorsement from Bertram Russell "Not since Euclid's Elements have we seen anything like it'. Others have dismissed it as a work of mysticism with little mathematical content. This view has not been mitigated by Spencer-Brown's subsequent claim to be the reincarnation of the Buddha. His work has largely been ignored by the mathematical community.

The major exception to this is Louis Kauffman, of the University of Illinois, who has made a major study of Spencer-Brown's work and has attempted to extend it. He may be looked to for a more balanced assessment. Kauffman concludes that Spencer-Brown has completed a program, begun by Peirce (2001):

> *Spencer-Brown's work can be seen as part of a continuous progression that began with Peirce's Existential Graphs. In essence what Spencer-Brown adds to the existential graphs is the use of the unmarked state. That is, he allows the use of empty space in place of a complex of Signs. This makes a profound difference and reveals a beautiful and simple calculus of indications underlying the existential graphs.*

Kauffman also makes available a draft of his book concerning Laws of Form (2011). This in-depth scholarly work contains much of the content of his earlier paper but also demonstrates the application of Spencer-Brown's paradigm to topics in biology, quantum physics and topology.

> *There is a kind of blinding clarity about these simple ideas near the beginning of Laws of Form. They point to a clear conception of world and organism arising from the idea of a distinction. Nevertheless, if you follow these ideas out into any given domain you will be confronted by, perhaps engulfed by, the detailed complexities of that domain.*

In brief Kauffman accepts Spencer-Brown's work as foundational and endorses his conclusions concerning the existence of a world mind; the striving of the universe to know itself. This conclusion is base on the understanding that at the bottom of existence is a distinction between a state which knows and a separate state which is known. However this understanding is mathematical and little guidance is offered as to its physical implementations.

A path I have outlined is to consider Spencer-Brown's distinction within the context of semiotics, as suggested by Kauffman, and to further consider Bayesian inference as a well developed mathematical description of semiotic systems. In this view distinction is seen as fundamental to Bayesian inference as it is the unit of information.

Further inferential systems are implemented within a wide range of physical settings as Darwinian processes. Darwinian processes are perhaps the fundamental mechanism within nature whereby one state may come to know another and are utilized by a plethora of scientific theories across a broad expanse of subject matters; as examples: cosmology (Smolin 1997), fundamentals of quantum theory (Zurek 2009), biology (Darwin 1872), neuroscience (Adams 1998) and culture (Blackmore 1999). The noting of the existence of this set of unified interdisciplinary scientific theories is the motivation for the meta-theory of universal Darwinism.

Universal Darwinism also strives to provide insight into the details of the Darwinian process. We might view the fact that each of these many theories describe information processing systems designed to increase knowledge as evidence for the existence of a world mind and that a fundamental aspect of the universe is its evolving ability to know itself.